\documentclass[pra,twocolumn,a4paper,showpacs,superscriptaddress,prl]{revtex4}
\usepackage{amsmath}
\usepackage{amsfonts}
\usepackage{graphicx}
\usepackage{longtable}
\begin{document}
\title{Collective multipole-like signatures of entanglement in symmetric $N$-qubit systems} 
\author{A. R. Usha Devi}
\email{arutth@rediffmail.com}
\author{ R. Prabhu}
\affiliation{Department of Physics, Bangalore University, 
Bangalore-560 056, India}
\author{A. K. Rajagopal}
\affiliation{Department of Computer Science and Center for Quantum Studies, George Mason 
University, Fairfax, VA 22030, USA and
Inspire Institute Inc., McLean, VA 22101, USA.} 

\date{\today}

\begin{abstract} 
A cogent theory of  {\em collective multipole-like quantum correlations} 
in symmetric multiqubit states is presented by employing 
SO(3) irreducible spherical tensor representation.   
An {\em arbitrary bipartite division} of this system leads to a family 
of inequalities to detect entanglement involving averages of these 
tensors expressed in terms of the total system angular momentum 
operator. Implications of this theory to the quantum nature of 
multipole-like correlations of all orders in the Dicke states are deduced. 
A selected set of examples illustrate these collective tests. Such tests  
detect entanglement in macroscopic 
atomic ensembles, where individual atoms are not accessible.
\end{abstract}
\maketitle
Correlated macroscopic atomic ensembles~\cite{ensemble} offer promising possibilities in low-noise 
spectroscopy~\cite{Wineland}, high precision interferometry~\cite{Yurke}, and in the implementation of quantum 
information protocols~\cite{tele}. Experimental characterization of entanglement - which is the key ingredient in    
these applications - has attracted considerable attention. Main difficulty in analyzing 
such composite systems with large number, $N$, of particles is the corresponding exponential 
size of the Hilbert space.  So, major effort is focused  on exploring inseparability status of 
special classes~\cite{Vol, Stockton} of quantum states, confined to  smaller subspaces of the Hilbert space 
due to  symmetry requirements. For example, a macroscopic atomic ensemble of {\scriptsize$N$} two level atoms is a  
collective system of {\scriptsize$N$} spin-{\scriptsize$\frac{1}{2}$} systems (qubits), in a 
{\scriptsize$2^N$} dimensional Hilbert space  {\scriptsize${\cal H}=(C^2)^{\otimes N}$}. However, when the dynamics 
of the atomic ensemble is governed by collective operations - which do not address the atoms individually - 
the atoms  within the system are completely symmetric with respect to interchange. 
This class of permutation symmetric states, labeled by total spin 
  {\scriptsize$J=\frac{N}{2}$} (maximal value in the addition of {\scriptsize$N$} spin-{\scriptsize$\frac{1}{2}$} 
angular momenta), 
is restricted~\cite{Stockton} to a  {\scriptsize$N+1$} dimensional subspace  
 {\scriptsize${\cal H}_{\rm Sym}={\rm Sym}\, (C^2)^{\otimes N}$} (here `\,Sym' denotes symmetrization).
The eigen states {\scriptsize$\{|J=N/2,\, M\rangle ;\  \, -N/2\leq M\leq N/2 \}$} of total angular momentum operator
{\scriptsize$\hat{\vec J}=\frac{1}{2}\, \sum_{i=1}^N\, \vec{\sigma}_i$},  where {\scriptsize$\vec\sigma_i$} denotes 
Pauli spin operator of 
the {\scriptsize$i^{\rm th}$} atom,  span the space.   A much simpler analysis of inseparability in symmetric  atomic 
ensembles becomes possible in this space. As individual atoms are not  accessible in the macroscopic ensemble, only 
collective measurements are feasible and any test of entanglement requiring  individual control of atoms cannot be 
implemented experimentally.  For example, spin squeezing~\cite{kit}, i.e., reduction of quantum fluctuations in one of 
the spin component orthogonal to the mean spin direction below the fundamental noise limit {\scriptsize$N/4$} is an 
important {\em collective signature} of entanglement in symmetric {\scriptsize$N$} qubit systems, a consequence of 
two-qubit pairwise entanglement~\cite{Sor}. Recently~\cite{ARU}, necessary and sufficient conditions for pairwise 
entanglement have been formulated in terms of negativity of intergroup covariance matrix. 
These are  equivalent to the generalized spin squeezing inequalities~\cite{Kor} 
(for two qubit entanglement) involving collective first and second order moments of total 
angular momentum operator. Further, genuine three particle entanglement in symmetric multiqubit systems is 
shown~\cite{Kor} to obey inequalities involving bulk observables up to third order in total 
angular momentum {\scriptsize$\hat{\vec J}.$} 

In this paper, a family of sufficient conditions to detect entanglement  of atoms in a 
macroscopic ensemble, through collective measurements on the system, is derived.  
These are formulated in terms of covariance matrix condition involving averages of
SO(3) irreducible tensor operators {\scriptsize$\hat\tau^K_Q(N)$} of rank {\scriptsize $K=1,2, \ldots 
N$}, constructed from the total angular momentum operator 
{\scriptsize$\hat{\vec J}$} and are identified via 
an arbitrary bipartite split of the symmetric states of {\scriptsize$N$} atoms. The physical significance 
of these operators are that they express 
 {\scriptsize$K=1$}, dipole-like; 
{\scriptsize $K=2$}, quadrupole-like 
correlations etc., among the multiqubits.
The advantages of this procedure are mainly two-fold: one, the simplicity  in dealing 
 with a large variety of correlations in multiqubit systems, and two, the entanglement conditions 
  expressed in terms of experimentally observable signatures associated with the 
 correlations among irreducible tensor operators.  
This elegant formalism   enlarges the scope of the covariance matrix condition beyond those given 
in~\cite{ARU, ARU2}. The significance of this new approach 
is substantiated through illustrative  examples.

The SO(3) spherical tensor operators  {\scriptsize$\hat{\tau}^{K}_{Q}(N)$} are constructed  such that~\cite{Var, fn}  
their matrix elements in the  basis {\scriptsize$\{\vert N/2,\, M\rangle\}$} are given by 
{\scriptsize $\langle N/2,\,M'\vert \hat{\tau}^{K}_{Q}(N) \vert N/2,\, M\rangle~=~\sqrt{2K+1}\, C(N/2\,K\, 
N/2;MQM'),$}
in terms of the Clebsch Gordan (CG) coefficients~\cite{Rose}.   
These irreducible tensors are  orthogonal,  
{\scriptsize
${\rm Tr} (\hat{\tau}^{K}_{Q}(N)\, \hat{\tau}^{K'}_{Q'}(N)^\dag)=(N+1)\,\delta_{K,\,K'} 
\delta_{Q,\, Q'}$,}   
and the set {\scriptsize$\{ \hat{\tau}^{K}_{Q}(N); \ K=0,1,2, \ldots N,\  -K\leq Q\leq K\}$} 
forms a linearly independent basis  of operators in the Hilbert space of spin {\scriptsize$J=N/2$}  states. 
Thus a useful representation for the density  operator of symmetric  
{\scriptsize $N$}-qubits is given in terms of these operators by
{\scriptsize\begin{equation}
\label{rhosymcol}
\hat{\rho}(N)=\frac{1}{(N+1)}\, \displaystyle\sum_{K=0}^{N}\displaystyle\sum_{Q=-K}^{K}\, 
\hat{\tau}^{K\dag}_{Q}(N)\,\,  t^{K}_Q(N). 
\end{equation}}
It is completely specified by {\scriptsize$(N+1)^2-1$} irreducible tensor moments,
{\scriptsize\begin{equation}
\label{tkq}
 t^{K}_Q(N)= {\rm Tr}[\hat{\rho}_{\rm sym}(N)\, \hat{\tau}^{K}_{Q}(N)]=(-1)^Q\, t^{K*}_{-Q}(N), 
\end{equation}}
with {\scriptsize$t^0_0(N)=1$} because {\scriptsize${\rm Tr}\hat{\rho}~=~1$}. 
An important  {\em composition law} appropriate for examining multipole correlations between 
two constituent symmetric parts of this {\scriptsize$N$} particle system, characterized by angular momenta 
{\scriptsize$j_1~=~N_1/2$,} and {\scriptsize$j_2=N_2/2$} (with {\scriptsize$N=N_1+N_2$)}, of the ensemble  
is constructed in terms of direct product of  spherical tensors  {\scriptsize$\tau^{\kappa}_{q}(N_1)~\otimes 
~\tau^{\kappa '}_{q'}(N_2):$}     
{\scriptsize
\begin{equation} 
\label{rho12}
\hat{\rho}(N_1,N_2)=\frac{1}{(N_1+1)(N_2+1)}\, 
\sum (\hat{\tau}^{\kappa\dag}_{q}(N_1)\otimes 
\hat{\tau}^{\kappa '\dag}_{q'}(N_2))\, t^{\kappa\kappa'}_{qq'}(N_1,N_2),     
\end{equation}}
where {\scriptsize $t^{\kappa\kappa '}_{qq'}(N_1,N_2)={\rm Tr}[\hat{\rho}(N_1,N_2)\, \hat{\tau}^\kappa_{q}(N_1)\otimes 
\hat{\tau}^{\kappa '}_{q'}(N_2)]=(-1)^{q+q'}\,  
t^{\kappa\kappa '*}_{-q\, -q'}(N_1,N_2);$} {\scriptsize$\kappa=0,1,\ldots, N_1,\ \kappa'=0,1,\ldots, N_2;$} 
{\scriptsize$ -\kappa~\leq~q\leq \kappa,\, 
-\kappa'~\leq~q'~\leq~\kappa.$}
Equations (\ref{rhosymcol}) and (\ref{rho12}) represent the {\scriptsize$N$}-qubit symmetric  
system in two equivalent ways and thus the tensor parameters appearing therein 
are related as will be shown presently. These relations form the 
central core of the theory presented here. 

By taking trace over  {\scriptsize$N_2$} 
particles from the composite system (using the representation (\ref{rho12})), 
we obtain the density matrix of {\scriptsize$N_1$} particles. We find that 
the tensor parameters {\scriptsize$t^\kappa_q(N_1)$} characterizing the {\scriptsize$N_1$} 
subsystem are given by,  
{\scriptsize\begin{equation}
\label{N1N12}
t^\kappa_q(N_1)={\rm Tr}\left(\hat{\rho}(N_1)\, \hat{\tau}^\kappa_{q}(N_1)\right) 
=t^{\kappa 0}_{q0}(N_1,N_2);\ \  \kappa=0,1,\ldots, N_1.
\end{equation}}
Similar consideration for {\scriptsize$N_2$}-subsystem leads to 
{\scriptsize$t^{\kappa '}_{q'}(N_2)=t^{0\kappa '}_{0q'}(N_1,N_2).$}
The second set of relations connecting the tensor parameters of the bipartite
system given by (\ref{rho12}) with those  of (\ref{rhosymcol}), 
are obtained by using the orthogonality 
property of the tensor operators: 
{\scriptsize\begin{equation} 
\label{tktktk}
t^{\kappa\kappa'}_{qq'}(N_1,N_2)= \sum_{K,Q}\, {\cal F}^{(\kappa \kappa')}_{qq';\,K Q}(N_1,N_2,N)\, t^{K}_Q(N),
\end{equation}} 
where {\scriptsize$|\kappa-\kappa'|\leq K\leq \kappa+\kappa'$}; 
{\scriptsize$-K\leq Q\leq K $}, and 
{\scriptsize${\cal F}^{(\kappa \kappa')}_{qq';\,K Q}(N_1,N_2,N)$} are found to be 
{\scriptsize\begin{eqnarray}
\label{fkq}
&{\cal F}^{(\kappa \kappa')}_{qq';\,K Q}(N_1,N_2,N)={\rm Tr}\,  \left((\hat{\tau}^\kappa_{q}(N_1)\otimes 
\hat{\tau}^{\kappa '}_{q'}(N_2)
\, \hat{\tau}^{K}_{Q}(N)\right)\nonumber \\
\ \ &=\left[\frac{N_1}{2}\right]\, \left[\frac{N_2}{2}\right]\, 
\left[\frac{N}{2}\right]\, [\kappa]\, [\kappa']\,  C(\kappa\, \kappa'\,  
K; q\, q'\, Q)\,  
\left\{ \begin{array}{ccc}  \frac{N_1}{2} & \frac{N_2}{2} & \frac{N}{2}\\ 
 \frac{N_1}{2} & \frac{N_2}{2} & \frac{N}{2}\\  
\kappa & \kappa' & K \end{array} \right\}.\nonumber \\
\end{eqnarray}} 
Here `$\{\ \ \}$' 
denotes the Wigner-$9j$ symbol~\cite{Var,Rose} and  
{\scriptsize$[a]=\sqrt{2a+1}$}. 
Using the properties of the 9j-symbol and the CG coefficients~\cite{Var} 
for the special values {\scriptsize$\kappa'=0, q'=0$} in (\ref{tktktk}) and (\ref{fkq}), 
we obtain
{\scriptsize\begin{equation}
\label{tkqN12}
t^{\kappa 0}_{q0}(N_1,N_2)={\cal P}_\kappa(N_1, N_2)\,  t^\kappa_q(N), 
\end{equation}} 
where  {\scriptsize${\cal P}_\kappa(N_1,N_2)=\frac{N_1!}{(N)!}\sqrt{\frac{(N_1+1)(N+\kappa+1)!(N-\kappa)!}
{(N+1)(N_1+\kappa+1)!(N_1-\kappa)!}}\ ;\ \kappa=0,1,2,\ldots N_1$.}
By replacing {\scriptsize $N_2\rightarrow N_2-N_1,$}  (where  {\scriptsize $N_2\geq N_1) $}  
in both sides of (\ref{tkqN12}) and using (\ref{N1N12}), we obtain an important equivalent relation 
{\scriptsize\begin{equation}
\label{N12N}
t^\kappa_{q}(N_1)={\cal P}_\kappa(N_1, N_2-N_1)\,  t^\kappa_q(N_2). 
\end{equation}}
The product tensor parameters {\scriptsize$t^{\kappa\kappa '}_{qq'}(N_1,N_2)$} 
of the bipartite system exhibit a similar relationship with the 
corresponding coefficients {\scriptsize$t^{\kappa\kappa '}_{qq'}(\kappa,\kappa')$} of the 
{\scriptsize$\kappa+\kappa'$} subsystem. 
This follows by expressing {\scriptsize$t^{K}_Q(N)$} of the composite system
in the RHS of (\ref{tktktk}) in terms of  {\scriptsize$\kappa+\kappa'$} subsystem parameters 
 {\scriptsize$t^{K}_Q(\kappa+\kappa')$} (obtained by substituting {\scriptsize$N_1=\kappa+\kappa', 
N_2=N-(\kappa+\kappa')$} in (\ref{N12N})). 
Alternately,  we can also relate {\scriptsize$t^{\kappa\kappa '}_{qq'}(N_1,N_2)$} to 
{\scriptsize$t^{K}_Q(\kappa+\kappa')$} by choosing {\scriptsize $N_1=\kappa,\, N_2=\kappa'$} in (\ref{tktktk}). 
Comparing the resulting equations for {\scriptsize$t^{\kappa\kappa '}_{qq'}(N_1,N_2)$} and 
 {\scriptsize$t^{\kappa\kappa '}_{qq'}(\kappa,\kappa')$}  and using explicit expressions for the associated 
9j-symbols~\cite{Var} 
  we obtain, after some algebraic manipulation, 
{\scriptsize\begin{eqnarray}
 \label{N1N2kk}
t^{\kappa\kappa '}_{qq'}(N_1,N_2)&=&f(N_1,\kappa)\ f(N_2,\kappa')\ t^{\kappa\kappa '}_{qq'}(\kappa,\kappa'), \\
 f(N_\alpha,\kappa)&=& \sqrt{\frac{(N_\alpha+1)(2\kappa+1)!}{(\kappa+1)(N_\alpha+\kappa+1)!(N_\alpha-\kappa)!}};\ \ \  
\alpha=1,2. \nonumber 
\end{eqnarray}}
Equations (\ref{tktktk}), (\ref{N12N}) and (\ref{N1N2kk}) prove to be 
significant in identifying   collective signatures of entanglement in 
symmetric atomic ensembles,  obtained through an analysis of 
the bipartite representation (\ref{rho12}).

Consider a set of {\scriptsize$2\, (2\kappa+1)$} operators 
{\scriptsize$\hat A^\kappa_q~=~\hat{\tau}^\kappa_{q}(N_1)\otimes \hat I_{N_2}$} and 
 {\scriptsize$\hat B^\kappa_q~=~ \hat I_{N_1} ~\otimes ~\hat{\tau}^\kappa_{q}(N_2)$} 
 (where {\scriptsize$\hat I_{N_i}=\hat\tau^0_0(N_i)$} 
 denotes the identity operator).  Arranging  them as a column {\scriptsize$\xi^{(\kappa)}$} 
 (corresponding row of operators 
 being {\scriptsize$\xi^{(\kappa)}=(\hat A^{\kappa\dag}_q, \hat B^{\kappa\dag}_q$}),  define the 
{\scriptsize$2\kappa$}th order covariance matrix  for the symmetric system  as  
{\scriptsize \begin{equation}
V^{(2\kappa)}_{ij}~=~\frac{1}{2}\, \left\{\Delta\xi^{(\kappa)}_i,\ \Delta\xi^{(\kappa)}_j\right\}, 
 \end{equation}}
 where {\scriptsize$\Delta\xi^{(\kappa )}=\xi^{(\kappa)}-\langle\xi^{(\kappa )}\rangle$} and 
 {\scriptsize$\{\Delta\xi^{(\kappa )}_i,\ \Delta\xi^{(\kappa )}_j\}=$}{\scriptsize$\Delta\xi^{(\kappa )}_i\, 
\Delta\xi^{(\kappa )}_j+\Delta\xi^{(\kappa )}_j\, \Delta\xi^{(\kappa )}_i$}.  {\scriptsize$V^{(2\kappa )}$} exhibits a  
{\scriptsize$(2\kappa+1)\times (2\kappa+1)$} block matrix form, 
{\scriptsize$V^{(2\kappa  )}~=~\left(\begin{array}{cc}A^{(2\kappa  )}(N_1) & C^{(2\kappa  )}(N_1,N_2)\cr C^{(2\kappa  
)\dag}(N_1,N_2) & B^{(2\kappa  )}(N_2)\end{array}  \right).$} 
The diagonal blocks  {\scriptsize $A^{(2\kappa  )}_{qq'}(N_1)~=~\frac{1}{2}\, 
\langle\{ \Delta \hat A^\kappa_q, \Delta \hat A^\kappa_{q'}\}\rangle$},  
{\scriptsize $  B^{(2\kappa  )}_{qq'}(N_2)~=~\frac{1}{2}\, 
\langle \{\Delta \hat B^\kappa_q, \Delta \hat B^\kappa_{q'}\}\rangle$} correspond to multipole correlations 
among the  intra-group tensors and the off-diagonal block 
{\scriptsize$C^{(2\kappa  )}_{qq'}(N_1,N_2)=\frac{1}{2}\, \langle\{ \Delta \hat A^\kappa_q, \Delta \hat 
B^\kappa_{q'}\}\rangle$} 
comprises of inter-group multipole correlations. Explicitly, 
{\scriptsize\begin{equation} 
\label{c2k} 
C^{(2\kappa  )}_{qq'}(N_1,N_2)=(-1)^{q'}\,  [ t^{\kappa\kappa}_{q\, -q'}(N_1,N_2)- t^{\kappa0}_{q0}(N_1,N_2)\, 
t^{0\kappa}_{0-q'}(N_1,N_2)]. 
\end{equation}}
Here we focus on the {\scriptsize$(2\kappa~+~1)~\times~ (2\kappa~+~1)~$} hermitian 
cross-correlation matrix {\scriptsize$C^{(2\kappa  )}$}  and prove the following theorem:  

\noindent{\bf Theorem }: 
The cross-correlation matrix {\scriptsize$C^{(2\kappa  )}(N_1,N_2)$} of a given rank {\scriptsize $\kappa$} 
associated with any partition {\scriptsize$(N_1,N_2)$} of an {\scriptsize $N$}-qubit symmetric system is necessarily 
positive semidefinite for all separable symmetric 
{\scriptsize $N$}-qubit bipartite states. The sign of the corresponding 
matrix {\scriptsize$C^{(2\kappa  )}(\kappa,\kappa  )$}, associated with 
the  $2\kappa$ atom reduced system with equal partition, suffices to determine that of 
{\scriptsize$C^{(2\kappa  )}(N_1,N_2)$}, irrespective of the partitioning.

\noindent{\bf Proof}: In the product representation (\ref{rho12}) with 
an arbitrary partition {\scriptsize $(N_1,\, N_2)$}, 
a separable symmetric {\scriptsize $N$}-qubit state has the following structure: 
{\scriptsize\begin{equation} 
\label{sepsym} 
\hat{\rho}_{\rm sep}=\displaystyle\sum_w p_w\, \hat{\rho}_w(N_1) \otimes \hat{\rho}_w(N_2),\  
\ 0\leq p_w\leq 1; \ \ \sum_w p_w=1
\end{equation}} 
where {\scriptsize$\hat{\rho}_w(N_i)$} denotes the density matrices of the subensemble of $N_i$ qubits 
and is expressible as (\ref{rhosymcol}), 
in terms of  {\scriptsize$\hat{\tau}^\kappa_{q}(\hat{N_i})$}:
{\scriptsize$\hat{\rho}_{w}(N_i)~=~\frac{1}{(N_i+1)}\, 
\displaystyle\sum_{\kappa=0}^{N_i}\displaystyle\sum_{q=-\kappa}^\kappa\, 
\hat{\tau}^{\kappa\dag}_{q}(N_i)\,\,  t^\kappa_q(N_i,w).$} 
In a separable symmetric state 
(\ref{sepsym}) we have, 
{\scriptsize\begin{eqnarray} 
\label{sepsym2} 
t^{\kappa\kappa}_{qq'}(N_1,N_2)&=& {\rm Tr}\, \left( \hat{\rho}_{\rm sep}\, [ (\hat{\tau}^\kappa_{q}(N_1)\otimes 
\hat{\tau}^\kappa_{q'}(N_2)]\right)\nonumber \\ 
&=&\displaystyle\sum_w p_w\, {\rm Tr}\, \left( \hat{\rho}_w(N_1)\, 
\hat{\tau}^\kappa_{q}(N_1)\right) 
 {\rm Tr}\, \left( \hat{\rho}_w(N_2)\, \hat{\tau}^\kappa_{q'}(N_2)\right)\nonumber \\
&=&\displaystyle\sum_w p_w\, t^\kappa_{q}(N_1,w)\, t^\kappa_{q'}(N_2,w). 
\end{eqnarray}} 
Without any loss of generality, we may assume that {\scriptsize$N_2~\geq~ N_1$}. 
Now, employing  (\ref{tkqN12}) we express  
{\scriptsize$t^\kappa_{q'}(N_2, w)=[{\cal P}_\kappa(N_1,N_2-N_1)]^{-1}\, t^\kappa_{q}(N_1,w)$},  
which leads to an interesting  form for the matrix 
{\scriptsize$C^{(2\kappa  )}(N_1,N_2)$} of (\ref{c2k}) in a separable symmetric state:  
{\scriptsize\begin{equation} 
\label{s2ksep}
C^{(2\kappa  )}_{q\, q'}(N_1,N_2)
=[{\cal P}_\kappa(N_1,N_2-N_1)]^{-1}\, C^{(2\kappa  )}(N_1,N_1) 
\end{equation}}
Consider an  hermitian quadratic form 
{\scriptsize\begin{equation} 
\label{qform}
Q^\kappa=X^{\kappa\dag}\, C^{(2\kappa  )}(N_1,N_2) X^\kappa  
= \sum_{q,\ q'=-\kappa}^\kappa\, C^{(2\kappa  )}_{q q'}(N_1,N_2)\, X^{\kappa*}_{q}\, X^\kappa_{q'},
\end{equation}} 
with  {\scriptsize$X^\kappa\in R^{(2\kappa+1)}$} being an arbitrary real column vector, whose spherical 
 components~\cite{Var} are denoted  by {\scriptsize$X^\kappa_q=(-1)^{q}\, X^{\kappa*}_{-q}$}. 
In a separable symmetric state (\ref{sepsym}) it is readily seen that   
{\scriptsize $  
Q^\kappa_{\rm sep}~=~[{\cal P}_\kappa(N_1,N_2~-~N_1)]^{-1}\, \Pi^\kappa_{\rm sep}\geq 0$}, 
with {\scriptsize$\Pi^\kappa_{\rm sep}=X^{\kappa\dag}\, C^{(2\kappa  )}(N_1,N_1) X^\kappa = 
\sum_w p_w [\sum_q\ X^{\kappa*}_{q}\,   
t^\kappa_{q}(N_1,\, w)]^2\, -[\sum_w p_w\, \sum_q\ 
X^{\kappa*}_{q}\,t^\kappa_{q}(N_1,\, w)]^2$}, a positive semidefinite quantity.  
This proves the first part of our theorem. 

The  second part of the theorem follows from  (\ref{N1N2kk}), leading to the result
{\scriptsize\begin{equation}
\label{CN1N2kk}
C^{(2\kappa  )}(N_1,N_2)=f(N_1,\kappa  )\, f(N_2,\kappa)\, C^{(2\kappa  )}(\kappa,\kappa  ),
\end{equation}} 
i.e., the covariance matrix {\scriptsize$C^{(2\kappa  )}(N_1,N_2)$} of an arbitrary symmetric system is proportional 
(with an overall positive multiplication factor) 
to that associated with the equal partitioning of  a $2\kappa$ qubit reduced system. 
(Note that the covariance matrix {\scriptsize$C^{(2\kappa  )}(N_1,N_2)$}
 could be related to its reduced system  counterpart  
{\scriptsize$C^{(2\kappa  )}(n_1,n_2)$}, which is then  seen to 
be proportional to {\scriptsize$C^{(2\kappa  )}(n_1,n_1)$}; {\scriptsize$n_1\leq n_2$}. Further,
{\scriptsize$C^{(2\kappa  )}(n_1,n_1)$} may be related to {\scriptsize$C^{(2\kappa  )}(n,n);\  n<n_1$} etc. 
This can go down all the way upto {\scriptsize$C^{(2\kappa  )}(\kappa,\kappa  )$}. Hence, 
the positivity (negativity) of  {\scriptsize$C^{(2\kappa  )}(N_1,N_2)$} 
has its origin in the {\scriptsize $2\kappa$}-qubit
 covariance matrix {\scriptsize$C^{(2\kappa  )}(\kappa,\kappa  )$}, with equal partition {\scriptsize$(\kappa,\kappa  
)$.})  $\Box$ 

Thus, for any arbitrary (pure or mixed) symmetric ensemble 
of  qubits,  {\scriptsize$C^{(2\kappa  )}(\kappa,\kappa  )<0,$} for various orders   
{\scriptsize$\kappa=1,2,\ldots \, ; 2\kappa~\leq N$}, serves as a sufficient condition 
of  entanglement and leads to a family of inseparability conditions 
associated with the  quantum correlations between inter-group tensor 
operators i.e., {\scriptsize$\langle \Delta\hat A^{(\kappa  )}_q\, \Delta\hat B^{(\kappa  )}_q\rangle$}.   
As the product tensor parameters 
{\scriptsize$t^{\kappa\kappa}_{qq'}(\kappa,\kappa  )$} and 
{\scriptsize$t^{\kappa0}_{q0}(\kappa,\kappa  ),\, t^{0\kappa}_{0q})(\kappa,\kappa  )$}  - specifying the covariance 
matrix 
{\scriptsize$C^{(2\kappa  )}(\kappa,\kappa  )$}  -  are given in terms of collective tensor moments 
{\scriptsize$t^{K}_Q(N)$}, (see (\ref{tktktk}), (\ref{N12N}) and (\ref{N1N2kk})), 
the negativity of the covariance matrix {\scriptsize$C^{(2\kappa  )}(\kappa,\kappa  )$} is readily expressed in terms 
of averages of  symmetrized homogeneous {\scriptsize$2\kappa^{\rm th}$} order polynomials~\cite{fn} 
of {\scriptsize$\hat {\vec J}$},   thus leading to a family of 
{\em multipole-like collective signatures} of entanglement.  
  
We now show that the  spin squeezing inequality~\cite{ARU, Kor} is a consequence of the intergroup dipole correlations
( for {\scriptsize$\kappa=1$}) viz.,  {\scriptsize$C^{(2)}(N_1,N_2)<0$}:  
In the {\scriptsize$3\times 3$} matrix {\scriptsize$C^{(2)}(N_1,N_2)$} we substitute 
{\scriptsize $t^{\kappa\kappa '}_{qq'}(N_1,N_2); \kappa,\kappa'=0,1$} in terms of total system 
collective parameters {\scriptsize$t^K_Q(N)$} (see (\ref{tktktk})) using explicit values~\cite{Var} 
for Wigner $9j$ symbols and CG coefficients. Following this by a unitary transformation 
corresponding to a change  from   
spherical basis~\cite{Var} {\scriptsize${\bf e}_\mu ,\ \mu=\pm 1, 0$} to Cartesian basis  
{\scriptsize${\bf e}_i ,\ i=x,y,z$} leads to  
{\scriptsize\begin{equation}
\label{genspsq1}
U\, C^{(2)}(N_1,N_2)\, U^\dag={\cal A}\,  
[-\frac{N}{4}\, {\cal I}+ V + \frac{1}{N}\, SS^T]
\end{equation}}   
(where {\scriptsize${\cal I}$} denotes the {\scriptsize$3\times 3$} identity matrix; 
{\scriptsize$V_{\alpha\, \beta}~=~\frac{1}{2}\, \left \langle ({\hat J}_{\alpha}{\hat J}_{\beta}+
{\hat J}_{\beta}{\hat J}_{\alpha})\right\rangle- \langle{\hat J}_{\alpha}\rangle\, 
\langle{\hat J}_{\beta}\rangle ;$ $S_{\alpha}=\langle{\hat J}_{\alpha}\rangle$} and 
{\scriptsize ${\cal A}=\frac{12}{N(N-1)}
\sqrt{\frac{N_1\,N_2}{(N_1+2)(N_2+2)}}$}). 
Hence, {\scriptsize$C^{(2)}(N_1,N_2)<0\Longleftrightarrow U\, C^{(2)}(N_1,N_2)\, U^\dag<0 \Longleftrightarrow 
V + \frac{1}{N}\, SS^T <\frac{N}{4},$} 
which is the known spin squeezing inequality~\cite{ARU, Kor}, deduced here from any bipartite division 
{\scriptsize$(N_1,N_2)$}. (The covariance matrix condition of \cite{ARU2} coincides with that given 
here {\em only} for  dipole correlations.)

An important application of our theorem concerns the Dicke states~\cite{Dicke} 
{\scriptsize$\vert \frac{N}{2}, M\rangle$; 
$ -\frac{N}{2}\leq M \leq \frac{N}{2}$}. These states 
provide an excellent set of physically relevant multiatom  symmetric states for  illustrating 
other types of multipole correlations. The collective tensor moments for the Dicke states 
  are   {\scriptsize$t^K_Q(N)=\langle \frac{N}{2}, M\vert 
\hat\tau^K_Q(N/2)\vert\frac{N}{2}, M\rangle=[\kappa]\, 
C(\frac{N}{2}\, K\, \frac{N}{2}; M\, 0\, M)\, \delta_{Q\, 0}$}. This corresponds to non-zero product tensor parameters 
{\scriptsize $t^{\kappa\kappa '}_{q-q}(N_1,N_2)$} (obtained from (\ref{tktktk})) implying that 
{\scriptsize$C^{(2\kappa  )}(N_1,N_2)$}  matrix   is diagonal  (see the definition (\ref{c2k})).  Except for  
{\scriptsize $\vert \frac{N}{2}, \pm\, \frac{N}{2}\rangle$}, which are product states, the covariance matrix 
{\scriptsize$C^{(2\kappa  )}(\kappa,\kappa  )$} 
of each of the Dicke states is negative, for all orders  {\scriptsize$\kappa$}, showing the 
quantum nature of multipole-like correlations of various orders.   
Recently~\cite{Koji}, an experimental scheme to reconstruct the spin-excitation number distribution 
of the collective spin states ( i.e., tomographic reconstruction of the diagonal elements of the 
density matrix in the Dicke basis) of macroscopic ensembles containing $\sim 10^{11}$ atoms, with low mean spin 
excitations, has been proposed. Implementation of such schemes would enable experimental detection of {\em collective 
quantum  multipole-like correlations} in macroscopic assembly  of entangled atoms. 

\begin{figure}
 \includegraphics*[width=1.82in,keepaspectratio]{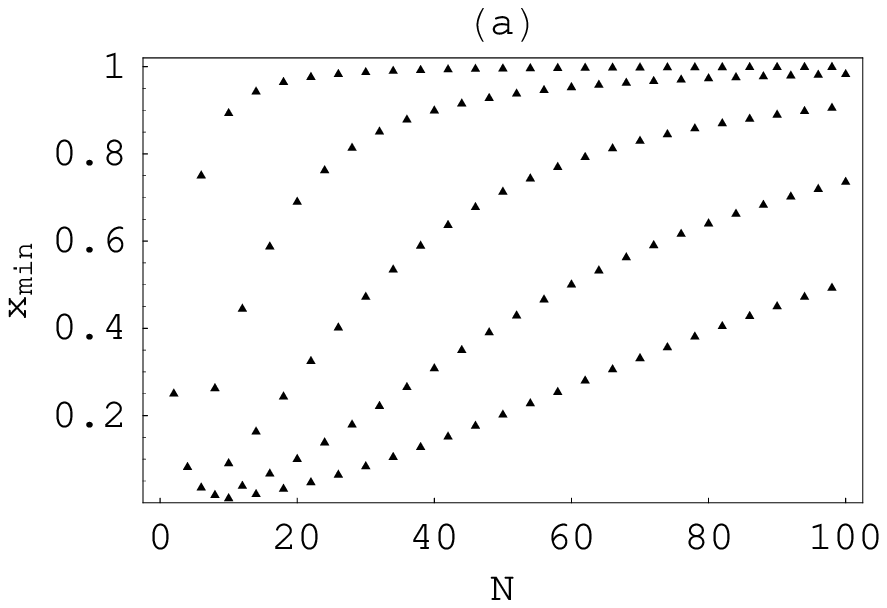}\includegraphics*[width=1.82in,keepaspectratio]{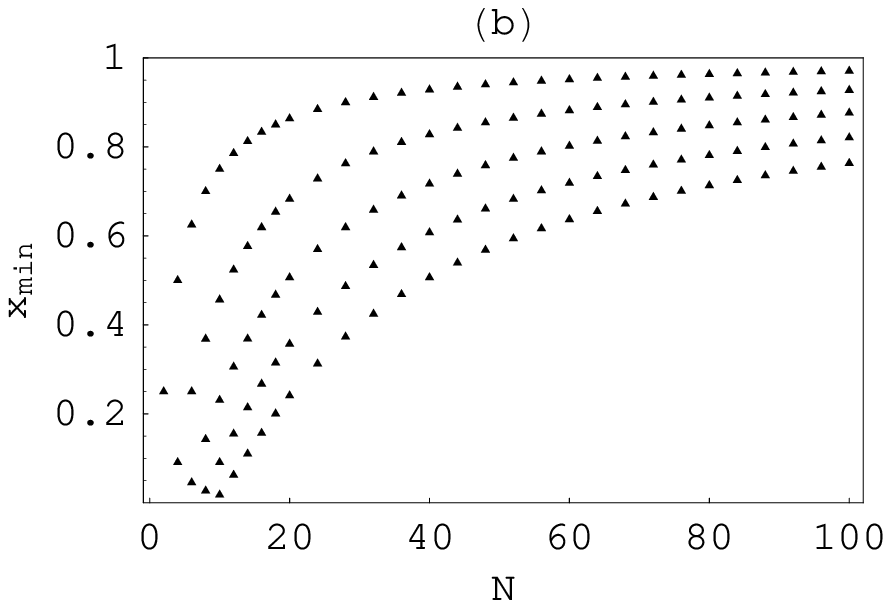}
 \includegraphics*[width=1.82in,keepaspectratio]{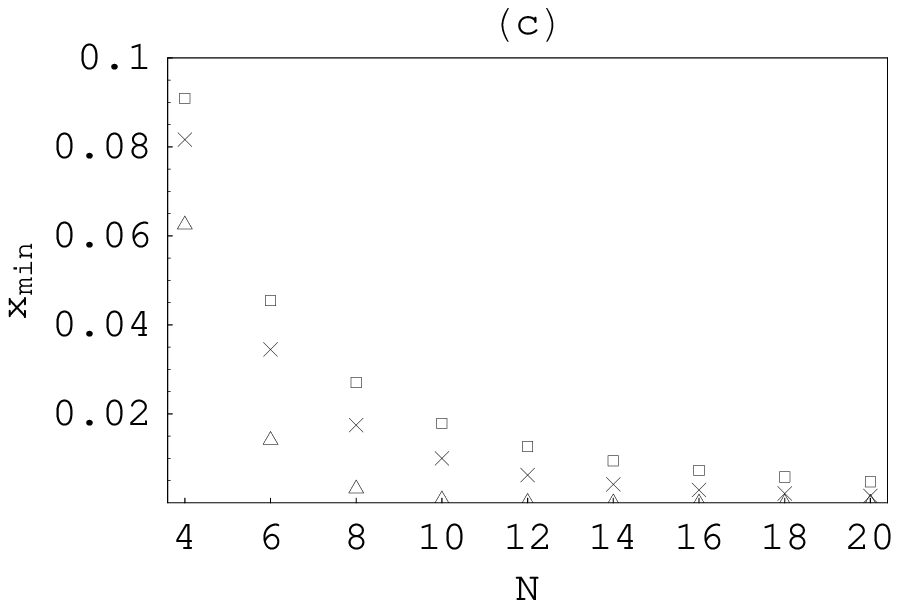}
 \caption{\scriptsize\label{fig1}Threshold value $x_{\rm min}$, evaluated from the multipole inseparability condition 
$C^{(2\kappa  )}(\kappa,\kappa  )<0$, with $\kappa=1$ to $\kappa=5$ (uppermost curve to the lower ones respectively) 
for the states  $\hat\rho_1$ (see (a))  and  $\hat\rho_2$ (see (b)),  as a function of the number of atoms $N$.
(c): $x_{\rm min}$ obtained from the negativity of the highest order covariance matrix $C^{(N)}(N/2,N/2)$ for the 
states $\hat\rho_1$ ($'\, +\,'$), $\hat\rho_2$ ({\scriptsize$'\,\Box\, '$}) and $\hat\rho_3$ ($\,'\times\,'$).}
\end{figure}

As an illustration of our method, we consider symmetric mixed states of the form 
{\scriptsize\begin{equation}
\label{noisy}
\hat{\rho}_i=\frac{(1-x)}{(N+1)}\, {\cal I}+x\, \vert\phi_i\rangle\, \langle \phi_i\vert;\ \  0\leq x\leq 1, 
\end{equation}}    
where {\scriptsize${\cal I}$} denotes  {\scriptsize$(N+1)\times(N+1)$} unit matrix; the states 
{\scriptsize$\vert\phi_i\rangle$}, for {\scriptsize$i=1,2,3$}, are given by   
{\scriptsize$\vert\phi_1\rangle=\vert \frac{N}{2},  
\frac{N}{2}-1\rangle$, $|\phi_2\rangle=\vert \frac{N}{2},  
0\rangle$} and {\scriptsize$\vert\phi_3\rangle=\frac{1}{\sqrt 2}(\vert \frac{N}{2},  
\frac{N}{2}\rangle+\vert \frac{N}{2},  
-\frac{N}{2}\rangle)$}.  We find the values of 
$x$, - using the condition {\scriptsize$C^{(2\kappa  )}(N_1,N_2)<0$} - as a function of  number of atoms, for  which 
{\scriptsize$\hat{\rho}_i$} of (\ref{noisy}) are inseparable. For the state {\scriptsize$\hat\rho_3$}  
we find that {\scriptsize$C^{(2\kappa  )}(\kappa,\kappa  )$},   are all positive for {\scriptsize$2\kappa<N$} and  
 the highest order co-variance matrix {\scriptsize$C^{(N)}(N/2,N/2)<0.$}
These results are presented in  graphical form in Fig.~\ref{fig1}. From Fig.~\ref{fig1}(a) and 
\ref{fig1}(b) it is clear that  dipole ({\scriptsize$\kappa=1$}) quantum correlations lead to 
{\scriptsize$x_{\rm min}\rightarrow 1$} for large 
{\scriptsize$N$} values, implying that the mixed states {\scriptsize$\hat\rho_i;\  i=1,2$} 
are separable throughout the  range {\scriptsize$0\leq x<1$} 
in this limit. However, higher order multipole correlations are more effective in revealing that these states are 
indeed entangled over a larger domain of  $x$.    The range of inseparability is sensitive to the difference 
{\scriptsize$N_e\sim N_g$} of the  number of atoms in ground and excited states ( which is {\scriptsize$N-1$} in  
{\scriptsize$\vert\phi_1\rangle$} and {\em zero} in {\scriptsize$\vert\phi_2\rangle$}). Further, from 
Fig.~\ref{fig1}(c) we find that highest order quantum correlations lead 
to    {\scriptsize $x_{\rm min}\rightarrow 0$} in the large {\scriptsize$N$} limit, implying that all the three mixed 
states {\scriptsize$\hat\rho_i$}, for {\scriptsize$i=1,2,3$}, are entangled in the range {\scriptsize$0< x\leq 1$}, 
when {\scriptsize$N\rightarrow\infty$}.
 
In conclusion, we have shown here that SO(3) irreducible tensor representation  
provides a powerful method to investigate inseparability in symmetric multiqubit 
systems. A family of sufficient conditions of inseparability
to  detect multipole-like {\em collective quantum correlations} derived here may be useful for experimental   
characterization of entanglement in  macroscopic 
atomic ensembles.  These  conditions are generalizations of the ones obtained in \cite{ARU, ARU2}. 
More specifically, instead of the Cartesian tensor product observables of Ref.~\cite{ARU2},  
 SO(3) irreducible tensor  observables are shown here to 
characterize any bipartite division {\scriptsize$(N_1,N_2)$} of symmetric 
{\scriptsize$N$}-qubit systems.  The two techniques 
 serve different purposes.  The 
first considers groups of qubits to  investigate the intergroup 
entanglement, whereas the second describes them in terms of
 spherical tensors involving collective {\scriptsize$N$}-qubit angular momenta. 
 The latter sheds light on entanglement among 
 dipole-like, etc.  multiqubit correlations,  
which may  be physically observable~\cite{Koji} in macroscopic ensembles of 
symmetric atoms.  An important consequence is 
that the Dicke states exhibit  quantum  
multipole-like correlations of all orders. Moreover, our approach is directly  applicable to characterize  
entanglement in 
spatially separated bipartite symmetric atomic ensembles, for example, two macroscopic gas samples of cesium 
atoms~\cite{Julsgaard}.  


\begin{thebibliography}{1}
\bibitem{fn}  The spherical tensors {\scriptsize$\hat{\tau}^{K}_{Q}$}  
are homogeneous polynomials of  order {\scriptsize$K$} constructed from the 
angular momentum operators {\scriptsize$\hat{\vec J}$} as,  
{\scriptsize$\hat{\tau}^{K}_{Q}= {\cal N}_{KJ}\,  (\hat{\vec J}\cdot \vec{\nabla})^K\ 
{\cal Y}_{KQ}(x, y, z)$} 
 where  {\scriptsize${\cal Y}_{KQ}(x, y, z)=r^K\, Y_{KQ}(\theta, \phi)$} are the solid harmonic functions. The 
  choice of normalization factor {\scriptsize${\cal N}_{KJ}~=~\frac{2^K}{K!}~\,
 \sqrt{\frac{4\pi\, (2J-K)!}{(2J+1)(2J+K+1)!}}$} leads to  
 their matrix elements in the {\scriptsize$\{\vert J\, M\rangle\}$} basis as given in  
the text. Note that the operators  {\scriptsize$\hat{\tau}^{K}_{Q}$}  
are related to the  {\scriptsize$\hat T_{KQ}$} of Ref.~\cite{Var} through  
{\scriptsize$\hat{\tau}^{K}_{Q}=[J]\, \hat T_{KQ}$}, so that 
{\scriptsize$\hat\tau^0_0$} is a {\scriptsize$(2J+1)\times (2J+1)$}  identity matrix.
\end{thebibliography}
\end{document}